\documentclass[10pt,aps,prd,twocolumn,showpacs,amsmath,amssymb,nofootinbib,eqsecnum,preprintnumbers,longbibliography,superscriptaddress]{revtex4-2}
\usepackage[english]{babel}				
\usepackage[mathscr]{eucal} 			
\usepackage{accents}					
\usepackage{mathtools}					
\usepackage[ddmmyyyy]{datetime}    		

\usepackage{graphicx}
\usepackage{xcolor}
\usepackage{float}

\newcommand\feq{\mathrel{\phantom{=}}}

\begin{document}
	
\title{Kerr-Schild perturbations in higher derivative gravity theories in $D$ dimensions}

\author{Ercan Kilicarslan}

\email{ercan.kilicarslan@usak.edu.tr}

\affiliation{Department of Mathematics, Usak University, 64200, Usak, Turkey.}

\date{\today}
	
\begin{abstract} 
We study Kerr-Schild perturbations of a ghost-free, generic non-local gravity theory constructed from an infinite series of higher derivative terms in $D$ dimensions. The infinite series of higher derivative terms are encoded by form factors, the forms of which can be restricted by requiring that the action remains perturbatively free of ghosts and tachyons around maximally symmetric backgrounds for transverse traceless fluctuations. To demonstrate this, we  obtain field equations for AdS plane wave metric in Kerr-Schild form, which yield linearized field equations for transverse-traceless spin-$2$ field. Using unitarity and consistency requirements, we obtain, as an example, the explicit derivation of non-local form factors in $D$ dimensions. 
\end{abstract}

\maketitle
\section{Introduction}
Einstein's general relativity (GR) has provided many successful observations and predictions, such as gravitational waves, black holes \cite{will}. However, it is not a complete theory at both large (IR regimes) and short (UV regimes) distance scales. At large distances, GR cannot account for the accelerated expansion of the universe and the rotational curves of outer objects in galaxies without invoking dark energy and dark matter. On the other hand, as for the short distances, it suffers from cosmological and black hole type spacetime singularities at the classical level \cite{Hawking}; at the quantum level, it is not a perturbatively renormalizable theory.

It has been recently shown that the modification of GR with infinite series of higher-derivative terms incorporating the non-locality has the potential to provide a well-defined theory in the short distances \cite{Efimov:1967pjn, Moffat:1990jj,Evens:1990wf,Tomboulis:2015gfa,Buoninfante:2018mre,Tomboulis:1980bs,Tomboulis:1997gg,Modesto1}. Non-local theories described by an action constructed from analytic form factors which give rise to non-local interactions. In particular, infinite derivative gravity (IDG) is free from the ghost like and black hole or cosmological type singularities, in which the propagator in a flat background is given by modification of the pure GR propagator via an exponential of an entire function which has no roots in the complex plane \cite{Biswas2,Biswas1}. This modification ensures that the theory is free from ghost-like instabilities and does not introduce any extra degrees of freedom (DOF) beyond the massless graviton. Moreover, the infinite derivative extension of Einstein's gravity leads to a non-singular Newtonian potential for a point-like source at short distances \cite{Biswas1,Edholm:2016}. This result is further extended to include cases where point-like sources also have velocities, spins, and orbital motion, leading to additional spin-spin and spin-orbit interactions alongside the usual mass-mass interactions \cite{Kilicarslan:2018yxd}. Recently, there has been further progress in finding exact solutions of IDG \cite{Kilicarslan:2019njc,Dengiz:2020xbu,Kolar:2021rfl,Kolar:2021uiu}. On the other hand, propagators in a D- dimensional AdS background were constructed in \cite{Kolar:2023mkw}.

In this paper, we study transverse-traceless perturbations of the ghost free infinite derivative gravity in $D$ dimensions. The presence of an infinite series of higher-derivative terms makes it more difficult to study the perturbative stability of the theory. In the literature, the usual method involves decomposing the metric field into its degrees of freedom \cite{Biswas:2016etb,SravanKumar:2019eqt,Kolar:2023mkw}; however, this approach requires lengthy and complex calculations. Here, we consider a D-dimensional AdS plane wave metric in the Kerr-Schild form and derive the corresponding field equations, which also serve as the linearized field equations for transverse traceless metric perturbations $h_{\mu\nu}=2H\lambda_\mu\lambda_{\nu} $. This allows us to study the unitarity conditions and obtain the explicit form of $D$ dimensional analytic form factors which are not known in the literature for generic dimensions.  

The paper is organised as follows: In Sec.~\ref{sc:IDG}, we provide a short review of the ghost-free infinite derivative gravity. In Sec.~\ref{sc:APW}, we calculate field equations of IDG for the AdS-plane wave metric described in Kerr-Schild form. In Sec.~\ref{sc:Unitarity}, we analyze the perturbative stability of the theory by constraining the form factors and provide an explicit example. In Sec.~\ref{sc:CON}, we conclude by summarizing our main results. We also provide a supplementary material in the appendices.

\section{Infinite Derivative Gravity} \label{sc:IDG}
At small scales, general relativity is likely to be replaced by a well-behaved effective theory containing infinite series of higher-derivative terms, which can be written in the most general quadratic form in curvature, as given by the Lagrangian density in \cite{Biswas2,Biswas1,Biswas:2016etb,Biswas:2016egy} \footnote{We use mostly positive metric signature, $(-,+,+,+,...)$.}

\begin{equation}\label{action}
\begin{aligned}
\mathcal{L} &= \frac{\sqrt{-g}}{16\pi G}\Big[ R-2\Lambda_0 +\alpha_c\big(R {\cal
	F}_1 (\square_s) R +  R_{\mu\nu} {\cal
	F}_2(\square_s) R^{\mu\nu}
	\\
&\feq+ C_{\mu\nu\rho\sigma} {\cal
	F}_3(\square_s) C^{\mu\nu\rho\sigma}\big)\Big],
\end{aligned}
\end{equation}
in which ${G=M^{-2}_p}$ is Newton's gravitational constant, $\Lambda_0$ is bare cosmological constant, R is the scalar curvature, $R_{\mu\nu}$ is the Ricci tensor, $C_{\mu\nu\rho\sigma}$ is the Weyl tensor, ${\square_s\equiv\square/M_s^{2}}$, and $\alpha_c={1}/{M_s^2}$, where dimensionful constant $M_s$ denotes the \textit{scale of non-locality} at which non-local interactions become significant. In the $\alpha_c \to 0$ (or $M_s \to \infty$) limit, the theory reduces to GR. The form factors ${\cal F}_i(\square_s)$, which are analytic functions of d'Alembert operator
${\square\equiv g_{\mu\nu}\nabla^\mu\nabla^\nu}$, are given as

\begin{equation}
{\cal F}_i(\square_s)\equiv\sum_{n=0}^{\infty}f_{i,n}\frac{\square^n}{M_s^{2n}}, 
\label{idfunc}
\end{equation} 
in which $f_{i,n}$ are dimensionless coefficients. The form factors lead to non-local gravitational interactions and play an important role in avoiding ghost-like instabilities. The source-free field equations of motion for the action \eqref{action} are provided in Appendix~\ref{ap:EOM}.

\section{AdS-plane wave spacetimes in IDG}\label{sc:APW}
The field equations of infinite derivative gravity are highly complicated \cite{Biswas:2013cha}; therefore, attempting to analyze the unitarity conditions to ensure perturbative stability around constant curvature backgrounds is an highly nontrivial task. To overcome this difficulty, we consider $D$ dimensional metric in Kerr-Schild form,\footnote{For a more detailed discussion of the properties of Kerr-Schild metrics, see \cite{Kerr-Schild,Classification,Gullu:2011sj,Gurses:2012db}. }
\begin{equation}
g_{\mu \nu}=\bar{g}_{\mu \nu}+2 H \lambda_\mu \lambda_\nu,
\label{KSForm}
\end{equation}
where $\bar{g}_{\mu \nu}$ is the AdS background metric, $\lambda_\mu$ is a non-expanding, non-twisting, and shear-free null vector, $H$ is a scalar function and the following relations hold 
\begin{equation}
\begin{aligned}
\lambda^{\mu}\lambda_\mu=0,\quad\nabla_{\mu}\lambda_\nu=\xi_{(\mu}\lambda_{\nu)},\quad\xi_{\mu}\lambda^\mu=0,\quad \lambda^\mu\partial_\mu H=0.
\label{AdSplanewave}
\end{aligned}
\end{equation}
For the Kerr-Schild ansatz, the Ricci tensor can be calculated as \cite{Gursespp,Gurses:2012db,Gurses:2013jua}
\begin{equation}
R_{\mu\nu}=-\frac{D-1}{\ell^2}g_{\mu\nu}+\lambda_{\mu}\lambda_{\nu}\mathcal{O}H,
\label{Ricci}
\end{equation}
where $\ell$ is the AdS radius and the $\mathcal{O}$ operator is defined as
\begin{equation}
\mathcal{O}\equiv-\left(\square+2\xi^{\mu}\partial_{\mu}+\frac{1}{2}\xi^{\mu}\xi_{\mu}-\frac{2(D-2)}{\ell^{2}}\right).
\end{equation}
Notice that the traceless Ricci tensor can be calculated by using (\ref{Ricci}) takes the form
\begin{equation}
S_{\mu\nu}=\lambda_{\mu}\lambda_{\nu}\mathcal{O}H,
\label{TraclessRicci}
\end{equation} 
which belongs to type N according to null alignment classification \cite{Coley1,Coley2}. Furthermore, the scalar curvature and scalar invariants are constant for AdS wave spacetimes, thanks to this, non-local term $R {\cal
	F}_1 (\square_s) R$  produces only a constant term for the field equations. On the other hand, the following useful relations for the action of the d'Alembert operator can be obtained as \cite{Gursespp}:
\begin{equation}\label{eq:formulas}
\begin{aligned}
&\square(\lambda_\mu\lambda_\nu H) =\bar{\square}(\lambda_\mu\lambda_\nu H)=-\lambda_\mu\lambda_\nu\bigg(\mathcal{O}+\frac{2}{\ell^2}\bigg)H
\\
&\square^{n}S_{\mu\nu} =\bar{\square}^{n}S_{\mu\nu}=\left(-1\right)^{n}\lambda_{\mu}\lambda_{\nu}\left(\mathcal{O}+\frac{2}{\ell^{2}}\right)^{\!\!n}\!\mathcal{O}H,
\end{aligned}
\end{equation}
where $\bar{\square}=\bar{g}^{\mu\nu}\bar{\nabla}_\mu\bar{\nabla}_{\nu}$ is the d'Alembert operator of AdS background. In the course of computations, one must use the following identity of higher-order derivative of the Weyl tensor (see Appendix \ref{ap:KSCEOM}):
\begin{equation}
\nabla_\mu\nabla_\nu\square^{n}C^{\mu\alpha\nu\beta}=\frac{D-3}{D-2}\Big(\square+\frac{2R(D-2)}{D(D-1)}\Big)^{\!n}\Big(\square-\frac{R}{D-1}\Big)S^{\alpha\beta}.
\label{Weylidentity}
\end{equation}
By using the remarkable algebraic properties obtained above, highly complicated field equations of the theory for the AdS wave metric reduce to a more manageable form (for details of the calculations, see Appendix \ref{ap:EOM} and \ref{ap:KSCEOM}) \footnote{One can check the result by comparing it quadratic curvature gravity with suitable choice of form factors ${\cal F}_{1}=f_{1,0}=\alpha/ \alpha_c$,${\cal F}_{2}=f_{2,0}=\beta/ \alpha_c$ and ${\cal F}_{3}=0$ \cite{Gursespp}.},
\begin{equation}
\begin{aligned}
&\bigg(\Lambda_0 {+}\frac{(D-1)(D-2)}{2\ell^2}+\frac{\alpha_c(4-D)}{2D}\Big(f_{1,0}{+}\frac{f_{2,0}}{D}\Big)R^2\bigg)g_{\mu\nu}\\&{+}
\bigg[1{+}\alpha_c\Big[\Big(2f_{1,0}{+}\frac{2f_{2,0}}{D}\Big)R{+}\Big(\bar\square{+}\frac{2}{\ell^2}  \Big){\cal F}_{2}(\bar\square_s)
\\&
+\frac{4(D-3)}{D-2}{\cal F}_{3}\Big(\bar\square_s{-}\frac{2(D-2)}{M_s^2\ell^2}\Big)\Big(\bar\square+\frac{D}{\ell^2}\Big)\Big]\bigg]S_{\mu\nu}=0.
\label{ppgeneq1}
\end{aligned}
\end{equation}
The trace part of the equation 
 \begin{equation}
 \Lambda_0=-\frac{(D-1)(D-2)}{2\ell^2}-\frac{\alpha_c(4-D)}{2D}\Big(f_{1,0}{+}\frac{f_{2,0}}{D}\Big)R^2,
 \end{equation}
that gives a relation between the effective cosmological constant and AdS radius. Note that in $D=4$ \eqref{ppgeneq1} reduces to the field equations for AdS plane waves \cite{Dengiz:2020xbu} \footnote{Also, in the limit ${\ell\to\infty}$, (\ref{ppgeneq1}) reduces to field equations for pp-waves on Minkowski background \cite{Kilicarslan:2019njc} in $D=4$ dimensions.}.  On the other side, the trace-free part of the non-local field equations takes the following form 
\begin{equation}
\begin{aligned}
&\bigg[1{+}\alpha_c\Big[-\frac{D(D-1)}{\ell^2}\Big(2f_{1,0}{+}\frac{2f_{2,0}}{D}\Big){+}\Big(\bar\square{+}\frac{2}{\ell^2}  \Big){\cal F}_{2}(\bar\square_s)
\\&
+\frac{4(D-3)}{D-2}{\cal F}_{3}\Big(\bar\square_s{-}\frac{2(D-2)}{M_s^2\ell^2}\Big)\Big(\bar\square+\frac{D}{\ell^2}\Big)\Big]\bigg]\Big(\bar\square+\frac{2}{\ell^2}\Big)\lambda_\mu\lambda_\nu H=0.
\label{AdSppgeneq}
\end{aligned}
\end{equation}
It is important here to observe that the $D$ dimensional non-local field equations for AdS plane waves given by \eqref{AdSppgeneq}, are identical to the linearized field equations corresponding to the Kerr-Schild metric perturbations ${h_{\mu\nu}=g_{\mu\nu}-\bar{g}_{\mu\nu}=2H\lambda_{\mu}\lambda_{\nu}}$ which represents transverse-traceless spin-$2$ tensor fluctuations. Hence, we can consider the field equations \eqref{AdSppgeneq} to discuss perturbative stability of the theory around constant curvature backgrounds.  Accordingly, the canonical action for \eqref{action} can be written as
\begin{equation}
\begin{aligned}
& \delta^2 S=\frac{1}{2}\int   \sqrt {-\bar{g}} d^D x h^{\mu\nu}\bigg(\bar\square+\frac{2}{\ell^2}\Big)\bigg[1{+}\alpha_c\Big[\Big(\bar\square{+}\frac{2}{\ell^2}  \Big){\cal F}_{2}(\bar\square_s)\\&
\hskip 2.2 cm +\frac{4(D-3)}{D-2}{\cal F}_{3}\Big(\bar\square_s{-}\frac{2(D-2)}{M_s^2\ell^2}\Big)\bigg(\bar\square+\frac{D}{\ell^2}\Big)\\&\hskip 2.2 cm +\frac{D(1-D)}{\ell^2}\Big(2f_{1,0}{+}\frac{2f_{2,0}}{D}\bigg)\Big]\bigg]h_{\mu\nu}.
\end{aligned}
\end{equation}
Observe that the $\bar\square=-\frac{2}{\ell^2}\equiv \frac{\bar{R}}{6}$ pole corresponds to the usual massless graviton mode for Einstein's gravity. In the Minkowski limit, the spin-$2$ propagator is
\begin{equation}
\Pi=\frac{i}{p^2\bigg[1-\alpha_c p^2\bigg({\cal F}_{2}(-p_s^2)+\frac{4(D-3)}{D-2}{\cal F}_{3}(-p_s^2)\bigg)\bigg]},
\end{equation}
which reduces to the result obtained in for $D=4$ \cite{Biswas:2016egy,SravanKumar:2019eqt}. Now we can study the perturbative stability of the theory.

\section{Unitarity and Consistency Conditions}\label{sc:Unitarity}
We first note that by unitarity, we mean absence of ghosts and tachyons in the linearized excitations. It is also important to emphasize that we expect the theory to behave well at small distances relative to GR, reduce to GR at large scales, and contain no additional degree of freedom other than massless spin-$2$ graviton. Accordingly, the required condition is to avoid ghost like instabilities and satisfy the previously mentioned properties, form factors should be chosen as analytic functions with no zeros in the complex domain. To guarantee that the theory has no ghosts on the AdS background, the following operator  
\begin{equation}
\begin{aligned}
&{\cal{O}}(\bar{\square}_s,\ell)=\bigg[1{+}\alpha_c\Big[-\frac{D(D-1)}{\ell^2}\Big(2f_{1,0}{+}\frac{2f_{2,0}}{D}\Big)\\&\hskip 1.9 cm {+}\Big(\bar\square{+}\frac{2}{\ell^2}  \Big){\cal F}_{2}(\bar\square_s)
+\frac{4(D-3)}{D-2}\\&\hskip 1.9 cm\times{\cal F}_{3}\Big(\bar\square_s{-}\frac{2(D-2)}{M_s^2\ell^2}\Big)\Big(\bar\square+\frac{D}{\ell^2}\Big)\Big]\bigg].
\end{aligned}
\end{equation}
must have not any poles which leads to
\begin{equation}
{\cal{O}}(\bar{\square}_s,\ell)=e^{\gamma (\bar{\square}_s)},
\end{equation}
where $\gamma$ is an entire function that has no zeros in the complex plane. This provides two important results: first, the theory does not have additional degrees of freedom other than a massless spin-$2$ graviton, thereby avoiding ghost like instabilities; second, the graviton propagator is enhanced by an exponential factor, which leads to improved behavior at high momenta, leading to improved convergence in loop integrals.  We now consider an explicit example by choosing at least one of the analytic form factors to be non-vanishing:  
${\cal F}_{1}={\cal F}_{2}=0, {\cal F}_{3}\neq 0$ \footnote{Let us note that we write the form factor ${\cal F}_{3}$ in analytic form by rearranging entire function $\gamma (\bar\square)$. }
\begin{equation}
{\cal F}_{3}(\bar\square_s)=\frac{D-2}{4(D-3)}\frac{e^{\gamma (\bar{\square}_s+\frac{3D-4)}{M_s^2\ell^2})}-1}{(\bar{\square}_s+\frac{3D-4}{M_s^2\ell^2})},
\end{equation}
which leads the following second order action 
\begin{equation}
\delta^2 S=\frac{1}{2}\int   \sqrt {-\bar{g}} d^D x h^{\mu\nu}\Big(\bar\square+\frac{2}{\ell^2}\Big)e^{\gamma (\bar{\square}_s+\frac{D}{M_s^2\ell^2})}h_{\mu\nu}.
\end{equation}
which contains the usual spin-$2$ graviton pole and exponential enhancement in the UV limit (at large momenta $k\gtrsim M_s$).

On the other hand, one can also consider other possibilities by allowing at least one of the analytic form factors to be non-vanishing. The procedure will be the same as in the previous case.  As another example, with ${\cal F}_{2}={\cal F}_{3}=0, {\cal F}_{1}\neq 0$, the condition $\Big(1-\frac{2 f_{1,0} D(D-1)}{\ell^2}\alpha_c\Big)> 0$ should be satisfied.

\section{Conclusions}\label{sc:CON}
In this paper, we studied the Kerr-Schild perturbations at the quadratic level of the action for parity-invariant, ghost-free IDG in $D$ dimensions, which includes an infinite number of derivatives, around maximally symmetric spaces. Since the field equations of IDG include an infinite number of covariant derivatives, studying of the consistency and unitarity conditions of the theory may seem hopeless. At this point, one can suggest analyzing perturbative stability through tensor perturbations; however this method requires lengthy computations for IDG. Instead we considered the $D$-dimensional AdS plane wave metric in Kerr-Schild form and obtained field equations that are equivalent to linearized field equations for transverse-traceless metric perturbations $h_{\mu\nu}=2H\lambda_\mu\lambda_\nu$ which yields $h=0$. 

  We have shown that when the operator ${\cal{O}}(\bar{\square}_s,\ell)$ is given as the exponential of an entire function that has no zeros, the theory is perturbatively ghost free around maximally symmetric backgrounds. In other words, the graviton propagator is ghost free and enhanced by an exponential factor, leading to improved behaviour compared to GR. Moreover, the only propagating degree of freedom is the massless spin-$2$ graviton. We have also given an explicit example, in the limit ${\cal F}_{1}={\cal F}_{2}=0, {\cal F}_{3}\neq 0$,that the theory includes a nonlocal Weyl term as well as cosmological Einstein terms. In this case we constructed the analytic form factor ${\cal F}_{3}$ and showed that the theory contains usual massless spin-$2$ graviton pole and exponential enhancement at high momenta.

\section{Acknowledgments}
 We would like to thank Bayram Tekin and Suat Dengiz for useful discussions and suggestions. We are also grateful to I.~Kolář and T.~Málek for their valuable suggestions and for drawing our attention to their related work. The works of E.K. is supported by the TUBITAK Grant No. 119F241 and the Outstanding Young Scientist Award of the Turkish Academy of Sciences (TUBA-GEBIP).

\newpage
\appendix
\section{Equations of motion of IDG} \label{ap:EOM}
The source free field equations for the action \eqref{action} can be given as \cite{Biswas:2013cha}
\begin{widetext}
\begin{equation}
\begin{aligned}
&G^{\alpha\beta}
{+}\Lambda_0 g^{\alpha\beta}
{+}\frac{\alpha_c}{2}\Big[4G^{\alpha\beta}{\cal	F}_{1}(\square)R
{+}g^{\alpha\beta}R{\cal F}_1(\square)R
{-}4\left(\nabla^{\alpha}\nabla^{\beta}{-}g^{\alpha\beta}\square\right){\cal F}_{1}(\square)R
{-}2\Omega_{1}^{\alpha\beta}
{+}g^{\alpha\beta}(\Omega_{1}{}_\rho^{\rho}{+}\bar{\Omega}_{1}) 
{+}4R^{\alpha}{}_{\nu}{\cal F}_2(\square)R^{\nu\beta}
\\
&{-}g^{\alpha\beta}R_{\nu}{}^{\mu}{\cal F}_{2}(\square)R_{\mu}{}^{\nu}
{-}4\nabla_{\nu}\nabla^{\beta}({\cal F}_{2}(\square)R^{\nu\alpha})
{+}2\square({\cal F}_{2}(\square)R^{\alpha\beta})
{+}2g^{\alpha\beta}\nabla_{\mu}\nabla_{\nu}({\cal F}_{2}(\square)R^{\mu\nu})
{-}2\Omega_{2}^{\alpha\beta}
{+}g^{\alpha\beta}(\Omega_{2}{}^{\rho}_{\rho}{+}\bar{\Omega}_{2})
{-}4\Delta_{2}^{\alpha\beta}
\\
&{-}g^{\alpha\beta}C^{\mu\nu\rho\sigma}{\cal	F}_{3}(\square)C_{\mu\nu\rho\sigma}
{+}4C^{\alpha}{}_{\mu\nu\sigma} {\cal F}_{3}(\square)C^{\beta\mu\nu\sigma}
{-}4(R_{\mu\nu}+2\nabla_{\mu}\nabla_{\nu})({\cal F}_{3}(\square)C^{\beta\mu\nu\alpha})
{-}2\Omega_{3}^{\alpha\beta}
{+}g^{\alpha\beta}(\Omega_{3}{}_{\gamma}^{\gamma}{+}\bar{\Omega}_{3}) -8\Delta_{3}^{\alpha\beta}\Big]=0,
\label{IDGfeqns}
\end{aligned}
\end{equation}
where the symmetric tensors are
\begin{equation}
\begin{gathered}
\begin{aligned}
\Omega_{1}^{\alpha\beta} &=\sum_{n=1}^{\infty}f_{1,n}\sum_{l=0}^{n-1}\nabla^{\alpha}R^{(l)}\nabla^{\beta}R^{(n-l-1)},
&
\bar{\Omega}_{1} &=\sum_{n=1}^{\infty}f_{1,n}\sum_{l=0}^{n-1}R^{(l)}R^{(n-l)},
\\
\Omega_{2}^{\alpha\beta} &=\sum_{n=1}^{\infty}f_{2,n}\sum_{l=0}^{n-1}R_{\nu}{}^{\mu;\alpha(l)}R_{\mu}{}^{\nu;\beta(n-l-1)},
&
\bar{\Omega}_{2} &=\sum_{n=1}^{\infty}f_{2,n}\sum_{l=0}^{n-1}R_{\nu}{}^{\mu(l)}R_{\mu}{}^{\nu(n-l)},
\\
\Omega_{3}^{\alpha\beta} &=\sum_{n=1}^{\infty}f_{3,n}\sum_{l=0}^{n-1}C^{\mu}{}_{\nu\rho\sigma}^{;\alpha(l)}C_{\mu}{}^{\nu\rho\sigma;\beta(n-l-1)},
&
\bar{\Omega}_{3} &=\sum_{n=1}^{\infty}f_{3,n}\sum_{l=0}^{n-1}C^{\mu}{}_{\nu\rho\sigma}^{(l)}C_{\mu}{}^{\nu\rho\sigma(n-l)},
\end{aligned}
\\
\begin{aligned}
\Delta_{2}^{\alpha\beta} &=\frac{1}{2}\sum_{n=1}^{\infty}f_{2,n}\sum_{l=0}^{n-1}[R_{\sigma}{}^{\nu(l)}R^{(\beta|\sigma|;\alpha)(n-l-1)}-R_{\sigma}{}^{\nu;(\alpha(l)}R^{\beta)\sigma(n-l-1)}]_{;\nu},
\\
\Delta_{3}^{\alpha\beta} &=\frac{1}{2}\sum_{n=1}^{\infty}f_{3,n}\sum_{l=0}^{n-1}[C^{\rho\nu}{}_{\sigma\mu}^{(l)}C_{\rho}{}^{(\beta|\sigma\mu|;\alpha)(n-l-1)}-C^{\rho\nu}{}_{\sigma\mu}{}^{;(\alpha(l)}C_{\rho}{}^{\beta)\sigma\mu(n-l-1)}]_{;\nu}.
\end{aligned}
\end{gathered}
\end{equation}
where we used the notation for a power of d'Alembert operator, ${\square^n X^{\alpha\dots}_{\beta\dots}\equiv X^{\alpha\dots}_{\beta\dots}{}^{(n)}}$.

\section{Tensors in the field equations for the Kerr-Schild metric } \label{ap:KSCEOM}
In this part, we provide some details of the calculations. For this purpose, let us first derive the identity \eqref{Weylidentity}. The identity follows by bringing the term $\square$ in front of the two outermost covariant derivatives
\begin{equation}
\nabla_\mu\nabla_\nu\square^{n}C^{\mu\alpha\nu\beta}=\Big(\square+\frac{2R(D-2)}{D(D-1)}\Big)\nabla_\mu\nabla_\nu\square^{n-1}C^{\mu\alpha\nu\beta}=\Big(\square+\frac{2R(D-2)}{D(D-1)}\Big)^{n}\nabla_\mu\nabla_\nu C^{\mu\alpha\nu\beta}.
\end{equation}
\end{widetext}
Next using the identity
\begin{equation}
\nabla_\mu\nabla_\nu C^{\mu\alpha\nu\beta}=\frac{D-3}{D-2}\Big(\square-\frac{R}{(D-1)}\Big)S^{\alpha\beta},
\end{equation}
we finally obtain the identity \eqref{Weylidentity}. 

We now present computational details of the nonzero terms in the field equations \eqref{IDGfeqns} for the Kerr–Schild metric ansatz, treating each term individually:
\begin{enumerate}
\item Using the constancy of the scalar curvature, the following term in the field equations can be computed as
\begin{equation}
G^{\alpha\beta}{\cal	F}_{1}(\square)R=\Big(S^{\alpha\beta}-\frac{D-2}{2D}g^{\alpha\beta}R\Big)f_{1,0}R.
\end{equation}
\item Once again, since the scalar curvature is constant, the following term in the field equations becomes
\begin{equation}
R{\cal	F}_{1}(\square)R=f_{1,0}R^2.
\end{equation}
\item By employing the definition of the traceless Ricci tensor $S_{\mu\nu}=R_{\mu\nu}-\frac{R}{D}g_{\mu\nu}$, the term can be calculated as
\begin{equation}
R^{\alpha}{}_{\nu}{\cal F}_2(\square)R^{\nu\beta}=\frac{R}{D}\Big({\cal F}_2(\square)S^{\alpha\beta}+f_{2,0}R^{\alpha\beta}\Big).
\end{equation}
\item Once again, using the definition of the traceless Ricci tensor, the term in the field equations takes the form
\begin{equation}
R_{\nu}{}^{\mu}{\cal F}_{2}(\square)R_{\mu}{}^{\nu}=\frac{R^2}{D}f_{2,0}.
\end{equation}
\item By applying the identity $\nabla_{\nu}\nabla^{\beta}(H \lambda^{\nu}\lambda^\alpha)=\frac{R}{D-1}H\lambda^{\alpha}\lambda^\beta$, the term can be simplified to
\begin{equation}
\nabla_{\nu}\nabla^{\beta}({\cal F}_{2}(\square)R^{\nu\alpha})=\frac{R}{D-1}{\cal F}_{2}(\square)S^{\alpha\beta}.
\end{equation}
\item Making use of the traceless tensor definition, the term in the field equations can be written as
\begin{equation}
\square({\cal F}_{2}(\square)R^{\alpha\beta})=\square({\cal F}_{2}(\square)S^{\alpha\beta}).
\end{equation}
\item Recalling the identity \eqref{Weylidentity} (proof provided above), the term in the field equations can be calculated as
\begin{equation}
\nabla_{\mu}\nabla_{\nu}{\cal F}_{3}(\square)C^{\beta\mu\nu\alpha}=\frac{D-3}{2-D}{\cal F}_{3}\Big(\square+\frac{2R(D-2)}{D(D-1)}\Big)\Big(\square-\frac{R}{(D-1)}\Big)S^{\alpha\beta}.
\end{equation}
\item From the definition of the traceless Ricci tensor, it follows that
\begin{equation}
G^{\alpha\beta}
{+}\Lambda_0 g^{\alpha\beta}=\Big(\Lambda_0 {+}\frac{(D-1)(D-2)}{2\ell^2}\Big)g^{\alpha\beta}.
\end{equation}
\end{enumerate}
Note that each term derived here can be used to obtain the general field equations of the theory as given in \eqref{ppgeneq1}.

\end{document}